\documentclass[12pt]{article}
\usepackage{setspace}
\usepackage{amsmath}
\usepackage{graphicx}
\usepackage{cases}
\usepackage{subfigure}
\usepackage{natbib}
\usepackage{psfrag}
\usepackage{amssymb}
\usepackage{color,graphics,epsfig}
\usepackage{appendix}
\usepackage{fancyhdr}
\usepackage{xr}

\def\be{\begin{equation}}
\def\ee{\end{equation}}
\def\bea{\begin{eqnarray}}
\def\eea{\end{eqnarray}}
\def\bml{\begin{mathletters}}
\def\eml{\end{mathletters}}

\newcommand{\tc}{\textcolor}

\begin{document}

\title{Modes of rapid polygenic adaptation}

\author{Kavita Jain$^{\dagger}$ and Wolfgang Stephan$^{\S,\ddagger}$\\
$^{\dagger}$ Theoretical Sciences Unit, \\Jawaharlal Nehru Centre for Advanced Scientific Research, \\Jakkur P.O., Bangalore 560064, India \\
$^{\S}$Natural History Museum Berlin, \\ Invalidenstr. 43, 10115 Berlin, Germany\\$^{\ddagger}$ Biocenter, University of Munich,\\ Grosshaderner Str. 2, 82152 Planegg, Germany}

\date{\today}

\maketitle



\noindent
Correspondence:
\texttt{jain@jncasr.ac.in}, \texttt{stephan@bio.lmu.de}

\clearpage

\noindent\textbf{Abstract:} Many experimental and field studies have shown that adaptation can occur very rapidly. Two qualitatively different modes of fast adaptation have been proposed: selective sweeps wherein large shifts in the allele frequencies occur at a few loci and evolution via small changes in the allele frequencies at many loci. While the first process has been thoroughly investigated within the framework of population genetics, the latter is based on quantitative genetics and is much less understood. Here we summarize results from our recent theoretical studies of a quantitative genetic model of polygenic adaptation that makes explicit reference to population genetics to bridge the gap between the two frameworks. Our key results are that polygenic adaptation may be a rapid process and can proceed via subtle or dramatic changes in the allele frequency depending on \tc{black}{the sizes of the phenotypic effects relative to a threshold value.} We also discuss how the signals of polygenic selection may be detected in the genome. While powerful methods are available to identify signatures of selective sweeps at loci controling quantitative traits, the development of statistical tests for detecting small shifts of allele frequencies at quantitative trait loci is still in its infancy.

\section*{Introduction}

Adaptation may occur very rapidly in response to changes that may be natural or due to human activity. Some recent examples include color variation in guppies \citep{Reznick:2009}, field mice \citep{Vignieri:2010} and peppered moth \citep{Cook:2012}; insecticide resistance in Drosophila \citep{Ffrench:2002}; {beak size changes} in Darwin's finches \citep{Grant:2008} and limb development in Anolis lizards \citep{Losos:2009}. The genetic architecture underlying these phenotypic traits ranges from few genes of major effect \citep{Hof:2011} to highly polygenic systems \citep{Linnen:2013,Lamichhaney:2012, Lamichhaney:2015}. 

Monogenic adaptation, in which one or few loci in a neutral or weakly selected background are under positive selection, has been of interest since the influential work of \citet{Smith:1974}. Here a single or several allele frequencies at a selected locus undergo a large shift, possibly sweeping away the neutral genetic variation. Such selective sweeps have been proposed as a predominant mode of rapid adaptation \citep{Messer:2013} although the precise nature of this mechanism is still a matter of debate \citep{Jensen:2014}. Theoretical studies of selective sweeps have long been carried out within the framework of population genetics (reviewed by \citet{Stephan:2016}) but these theories do not say what happens at the phenotypic level, {\it i.e.}, how the change in the allele frequency translates into the evolution of the phenotype. 

In contrast, polygenic adaptation that involves a large number of selected loci has been traditionally studied using quantitative genetics \citep{Mackay:2004} which addresses the response of a quantitative trait to selection.  \tc{black}{Because} the quantitative genetic models date back to the time before the genetic mechanisms of inheritance were discovered, they do not make a reference to the underlying molecular details or dynamics.  However, some verbal arguments predict the allele frequencies to change by small amounts when a large number of minor genetic loci control a phenotypic trait  \citep{Pritchard:2010}. Yet, it is not clear if adaptation can occur rapidly via such subtle changes in the allele frequencies.

Thus, there has been a general disconnect between the theories of adaptation  that work at either phenotypic or genotypic level. 
Ideally one would like to consider models in which selection acts on the phenotypic trait 
which is connected to the underlying genetics through a genotype-phenotype map. The response to selection is then  found at the genetic level and predictions are made about phenotypic trait evolution. Such a roadmap has been developed by several workers including Bulmer \citep{Bulmer:1972}, Barton and Turelli \citep{Barton:1989b} and B{\"u}rger \citep{Burger:2000}, and we follow this direction here to understand the {rapid} evolutionary dynamics of a single quantitative trait under stabilizing selection \citep{Jain:2015,Jain:2017}.

Recently \citet{Pritchard:2010} have advanced the proposition that adaptation does not need to proceed via selective sweeps alone \citep{Messer:2013} and that it may involve modest changes in allele frequencies at many loci. Our analyses are in agreement with their proposal when the sizes of the effects contributing to a polygenic trait are small \tc{black}{relative to a threshold value (defined later).} However, we also point out that their perspective should be {enlarged} to include  selective sweeps as allele frequencies are found to undergo large shifts when effects are \tc{black}{larger than the threshold effect} \citep{Jain:2017}. 

\section*{Response of a quantitative trait after a sudden environmental shift}

Although the multilocus population genetics of quantitative traits has been of interest for a long time \citep{Burger:2000}, 
analytical results have been hard to come by \tc{black}{since} the relevant equations do not close (for e.g., the equation for the trait mean may  involve the trait variance whose evolution is determined by \tc{black}{trait skewness}, and so on). To overcome this technical difficulty, several different strategies have been employed: \tc{black}{when the phenotypic trait variance changes slowly, one may treat it as constant in time and thereby close the hierarchy of equations mentioned above \citep{Lande:1983,Chevin:2008,Jain:2015}. But such an approximation clearly fails when the trait variance changes rapidly \citep{Jain:2017}. }Another approach has been to {devise simple models} that describe specific situations \citep{Chevin:2008} but such models are either not general or detailed enough. Extensive numerical simulations of detailed models have been carried out \citep{Pavlidis:2012} but available computational power limits \tc{black}{such analyses to only a few loci}. 

Recently we analyzed a detailed quantitative genetic model that captures the \tc{black}{response of a polygenic trait subject to} stabilizing selection and mutation after a sudden shift in the phenotypic optimum \citep{Jain:2015,Jain:2017}. We considered a single trait that is determined additively (no dominance or epistasis) by $\ell$ diallelic loci in a large population of diploids. If the phenotypic effect of the $+$ allele at locus $i$  is $\gamma_i/2$ and the corresponding allele frequency is $p_i$, the mean phenotype $c_1$ and the genetic variance $c_2$ are given by \citep{Burger:2000}
\bea
c_1 &=& \sum_{i=1}^\ell \gamma_i (2 p_i- 1) \label{meandefn} ~, \\
c_2 &=& 2 \sum_{i=1}^\ell \gamma_i^2 p_i q_i \label{vardefn}~,
\eea
where $q_i=1-p_i$ is the frequency of the $-$ allele with effect $-\gamma_i/2$. The trait effects can be chosen from a gamma distribution \citep{Jain:2015} as it describes the livestock data on quantitative trait loci quite well \citep{Hayes:2001}. But, for simplicity, here we will assume that the effect size distribution is an exponential function with mean  ${\bar \gamma}$. 
We also assume that the fitness of an individual with trait value $z$ follows a Gaussian distribution centered about the phenotypic optimum $z'$, $w(z)=e^{-(s/2) (z-z')^2}$
where $s$ measures the strength of stabilizing selection on the trait. Then, in an infinitely large, randomly mating population, the change in the allele frequency at the $i$th locus due to selection and symmetric mutation is given by \citep{Barton:1986}
\be
{\dot p}_i= -s \gamma_i p_i q_i  (c_1-z')-\frac{s \gamma_i^2}{2} p_i q_i(q_i-p_i) +\mu (q_i-p_i)
~,~i=1,...,\ell ~,
\label{pevoleqnf}
\ee
where dot denotes the derivative with respect to time and $\mu$ is the mutation rate. 
The model defined by equation (\ref{pevoleqnf}) can be derived from the well known symmetric viability model \citep{Burger:2000} under loose linkage \citep{Jain:2017}. {The first term on the right-hand side of equation (\ref{pevoleqnf}) models directional selection toward the phenotypic optimum, the second term stabilizing selection in the vicinity of the optimum \citep{Wright:1935}, and the last term accounts for mutations \citep{Latter:1970,Bulmer:1972,Barton:1986}.}

\begin{figure} 
\begin{center}  
\includegraphics[width=0.8\textwidth,angle=270,keepaspectratio=true]{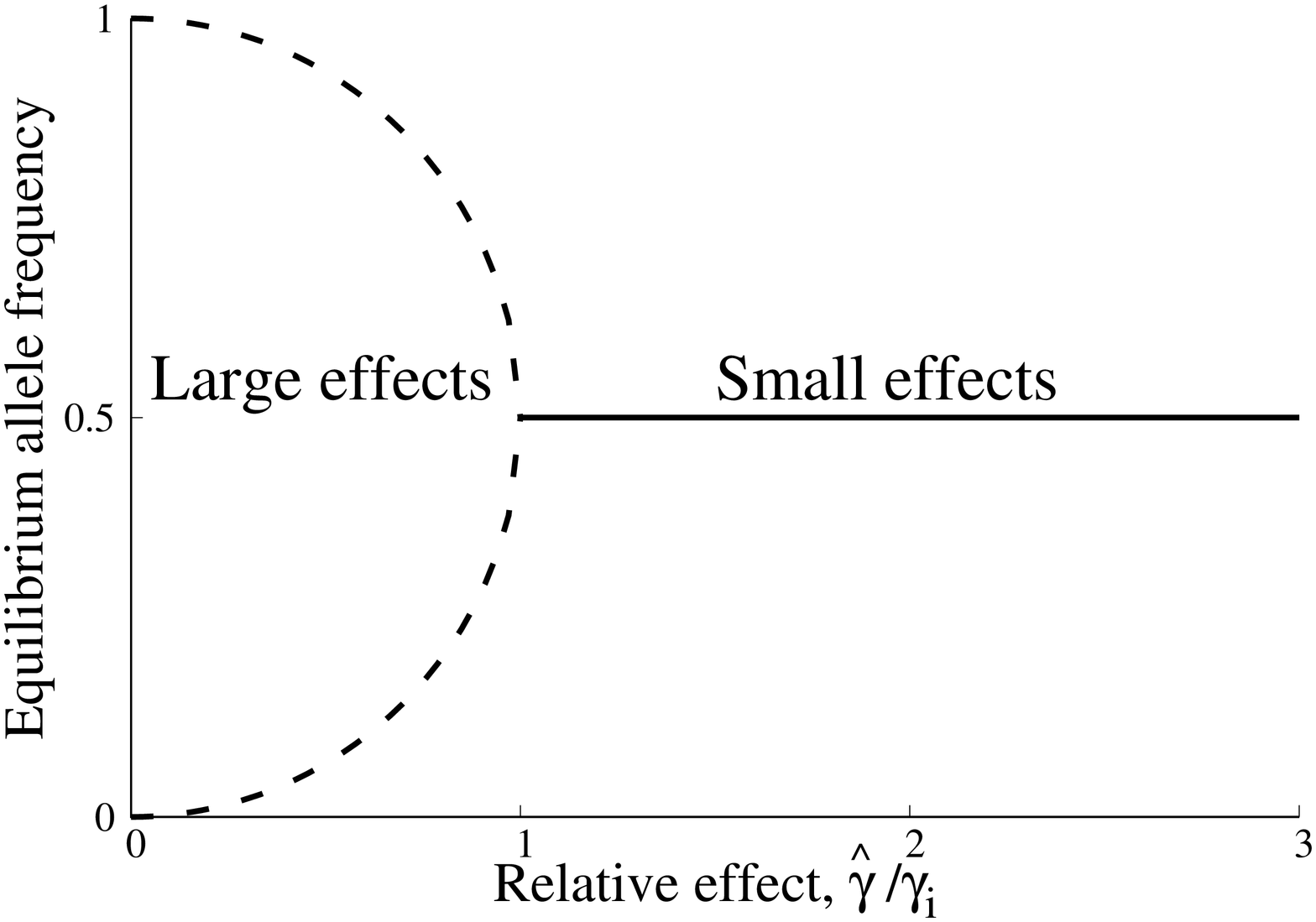} 
\end{center}
\caption{Relationship between the equilibrium allele frequency and the size of the effect. When the effect size is smaller than a threshold ${\hat \gamma}$, the allele frequency is one half but for larger effect size, two stable equilibria with allele frequency away from one half exist. These results hold when the stationary mean deviation is zero; the stable equilibria for nonzero stationary mean deviation are analyzed in \citet{Vladar:2014}.}
\label{fig_stat}
\end{figure}

Recently, \citet{Vladar:2014} presented an analytical treatment of the equilibrium properties of the above model and performed extensive numerical calculations. They found that the alleles may be classified into those with effects smaller than a threshold value ${\hat \gamma}=\sqrt{8 \mu/s}$ and those with larger sizes. This result is illustrated in Fig.~\ref{fig_stat} when the phenotypic mean coincides with the optimum, and we see that the equilibrium frequency of the alleles of small effect is one half, whereas the large-effect alleles are in a mutation-selection balance near zero or one when the effect size is much larger than the threshold effect. \tc{black}{This behavior of the allele frequencies} implies that when the mutation rate is sufficiently large, the stationary genetic variance given by equation (\ref{vardefn}) is also large while for sufficiently small mutation rates, it is negligible.

{A class of models in which a quantitative trait is controlled by a single locus with an infinite number of alleles   (continuum-of-alleles) and under stabilizing selection has also been investigated \citep{Burger:2000}. When the  phenotypic optimum coincides with the trait mean, as in the model under consideration, the stationary genetic variance is found to depend on whether mutation rate is large  (Gaussian model) \citep{Kimura:1965}) or small (House-of-Cards model)  \citep{Turelli:1984}) compared to the effect size. For the diallelic loci model described here, the variance per locus equals that of the House-of-Cards model when all loci are assumed to be of large effect but is smaller when a fraction of the loci has  small effect \citep{Vladar:2014,Jain:2015}.}

{We also note that in the model defined by equation (\ref{pevoleqnf}), the stationary genetic variance $c_2$ increases linearly with the number of loci irrespective of whether the effect size is small or large \citep{Vladar:2014}. This behavior is different from that in the infinitesimal model \citep{Bulmer:1980,Hill:2014,Barton:2017} in which a large number of loci each with small effect size (that decreases as $\ell^{-1/2}$) determines a quantitative trait resulting in a stationary genetic variance that is independent of the number of loci. More importantly, the infinitesimally small size of the effect results in a negligible change in allele frequency and variance for a finite change in the trait mean (${\dot p} \sim {\dot c_1} \ell^{-1/2}, {\dot c_2} \sim {\dot p}$ using equations (\ref{meandefn}) and (\ref{vardefn}), respectively). These properties then allow one to make predictions about the response to selection using the knowledge of variance in the base population (e.g., using breeder's equation)  at very short times \citep{Bulmer:1980,Hill:2014}. In contrast, here we assume that the effect size does not change with the number of loci and the variance can change substantially (see below).}

\begin{figure} 
\begin{center}  
\includegraphics[width=0.8\textwidth,angle=0,keepaspectratio=true]{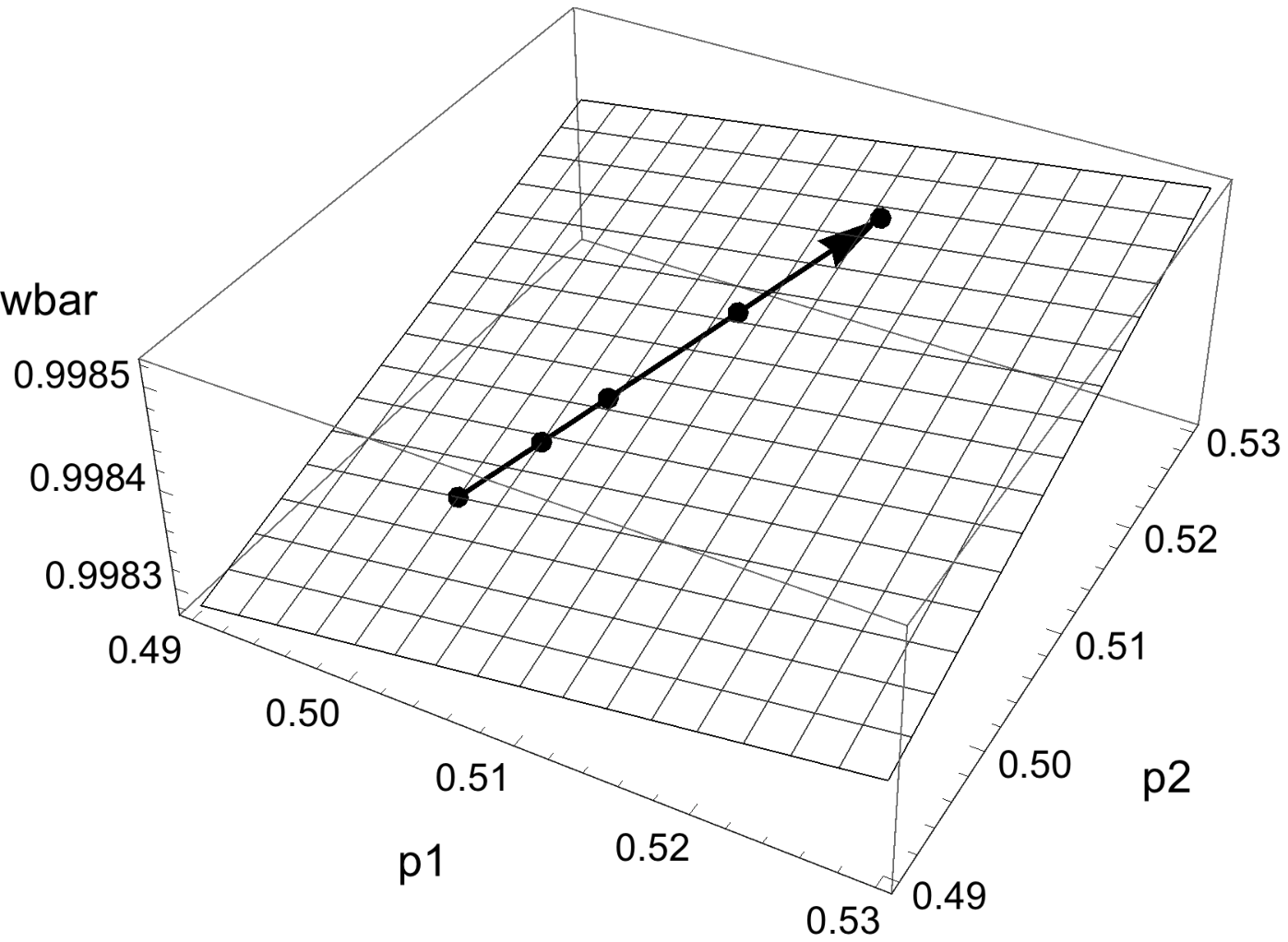}        
\includegraphics[width=0.8\textwidth,angle=0,keepaspectratio=true]{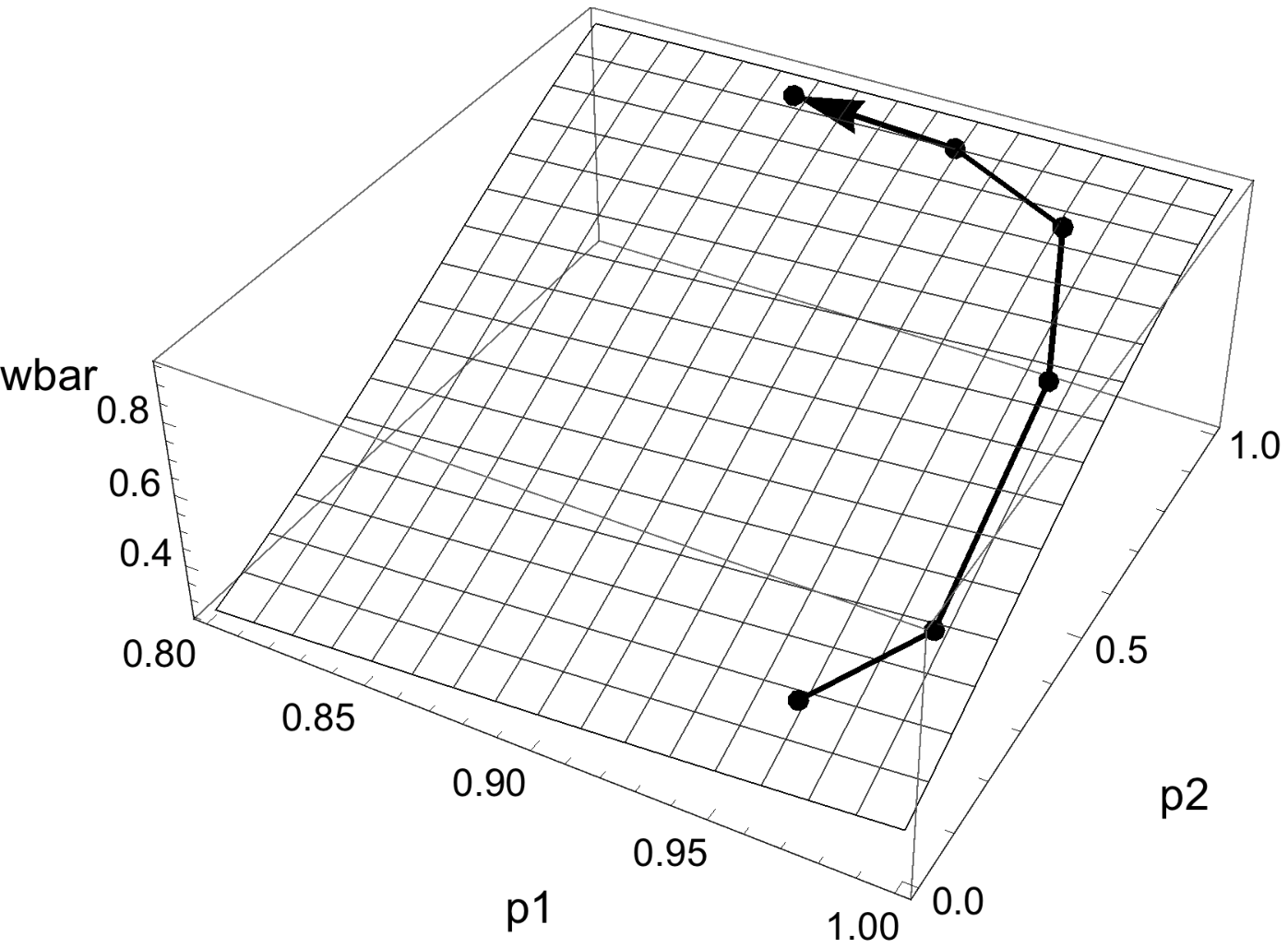} 
\end{center}
\caption{Evolution of allele frequencies on the adaptive landscape when a phenotypic trait is controlled by two loci. In the top (bottom) panel, both effects are smaller (larger) than the threshold effect. The points show the average fitness ${\overline w}=e^{-(s/2) \left[c_2+ (c_1-z_f)^2 \right]}$  at representative time points starting from a population equilibrated to a phenotypic optimum at zero until it reaches a stationary state at the new value $z_f$. The final average fitness is smaller than one as the genetic variance and the distance from the phenotypic optimum are nonzero in the new stationary state.} 
\label{fig_adaptive}
\end{figure}

\begin{figure} 
\begin{center}  
\includegraphics[width=0.8\textwidth,angle=0,keepaspectratio=true]{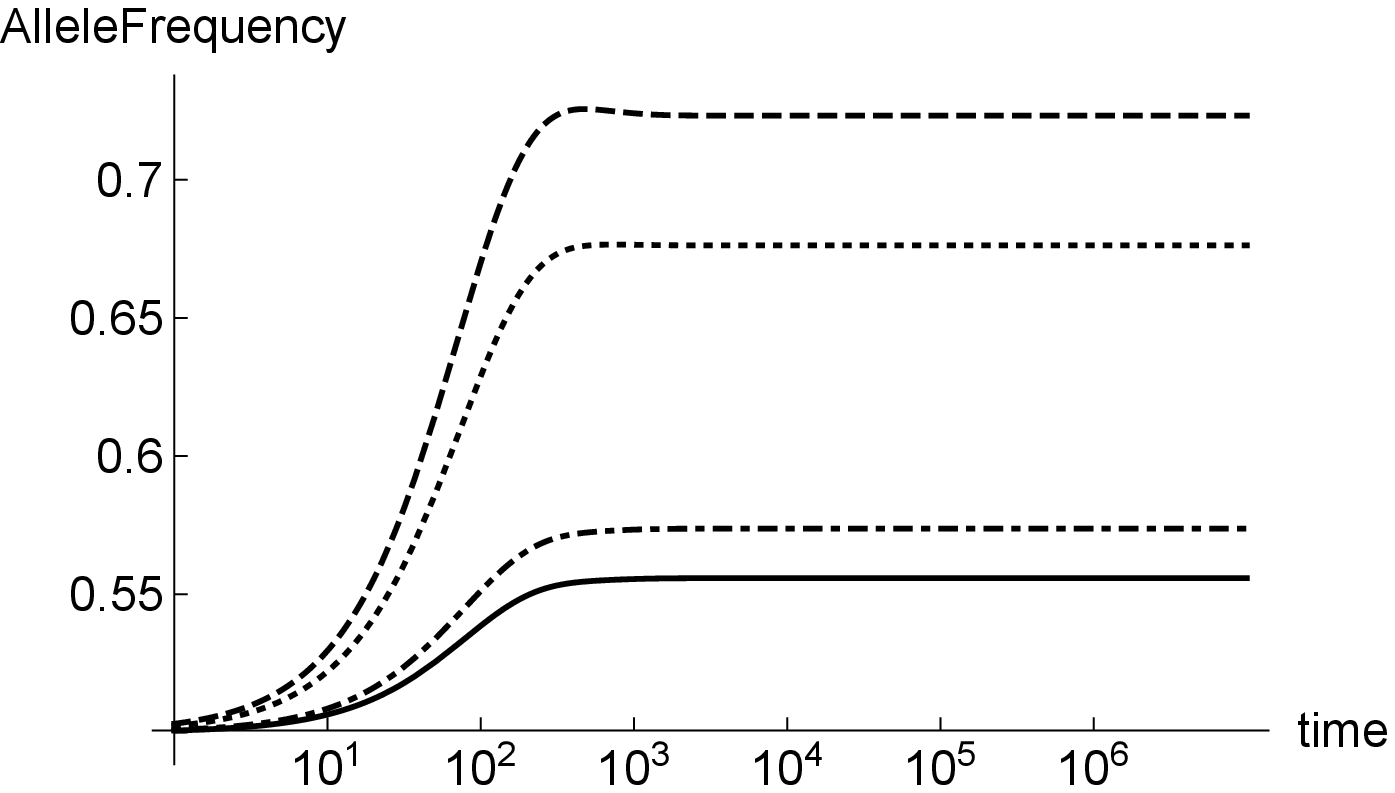}        
\includegraphics[width=0.8\textwidth,angle=0,keepaspectratio=true]{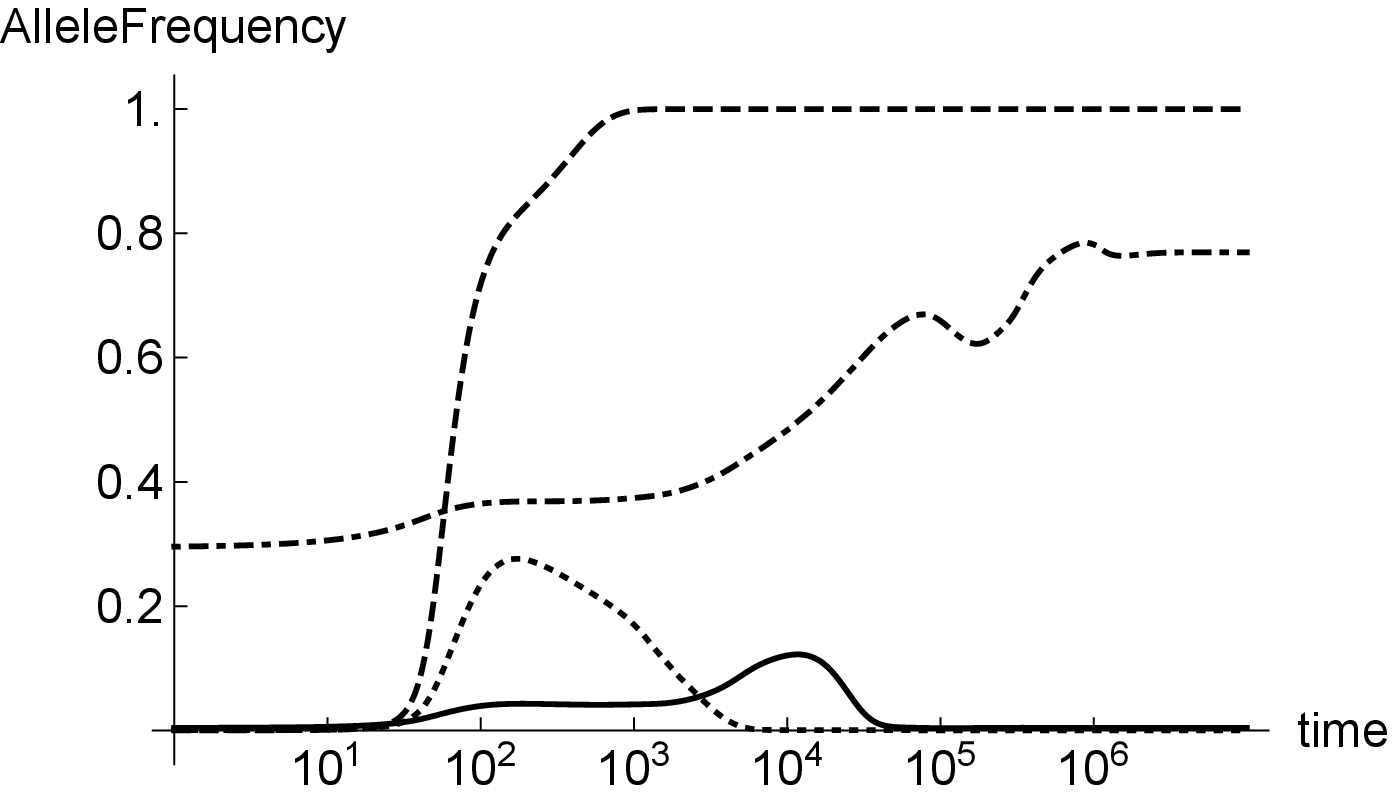}        
\end{center}
\caption{Allele frequency dynamics at four representative loci when most effects are small (top) and large (bottom) for $\ell=200$ loci. The sets of effects chosen from an exponential distribution (with mean ${\bar \gamma}=0.1$) are same in both graphs but the threshold effect is different. Note that in the large-effect case, some frequencies reach a stationary state at around $10^3-10^4$ time steps (short-term phase) while others take much longer.}
\label{fig_freq}
\end{figure}

To address the question of polygenic adaptation dynamics, we assume that the population is in equilibrium with no deviation from the phenotypic optimum located at $z_0$.  
Then, to describe fast evolution, the optimum is suddenly shifted to another value $z_f$ and in response, the allele frequencies evolve in time to the new stationary state. As an example, Fig.~\ref{fig_adaptive} shows the time evolution of the allele frequencies on the adaptive landscape \citep{Gavrilets:2004} when a trait is controlled by two loci. When both  effects are small (top panel), adaptation proceeds via small changes in the allele frequencies at both loci whereas a selective sweep occurs at the second  locus when both effect sizes are large (bottom panel). These qualitative features in the {dynamics of allele frequencies} - small to moderate changes at minor loci and selective sweeps at a few major loci that satisfy certain criteria (see the following section) - remain even when the number of loci is large as shown in Fig.~\ref{fig_freq} \citep{Jain:2017}. 

To calculate the response of the system after the optimum shift for large $\ell$, we now focus on the {\it short-term phase} which is defined as the one during which the mean reaches a value close to the new phenotypic optimum. During this phase, the full model defined by  equation (\ref{pevoleqnf}) can be approximated as \citep{Jain:2015, Jain:2017}
\be
{\dot p}_i= S_i p_i q_i ~,~i=1,...,\ell ~,
\label{dirsel}
\ee
where $S_i=-s \gamma_i (c_1-z_f)$. In the following, we call the model defined by equation (\ref{dirsel}) the {\it directional selection model}. In contrast to the classical model of directional selection \citep{Charlesworth:2010}, here the strength of selection also depends on the distance from the new phenotypic optimum. Moreover, because the mean deviation $c_1-z_f$ contains a sum over all allele frequencies, equation (\ref{dirsel}) represents a system of coupled nonlinear ordinary differential equations (ODEs) that are, in general, difficult to solve.  However, as shown in \citet{Jain:2017}, it is possible to obtain simple analytical expressions for the quantities of interest using the directional selection model {defined by equation (\ref{dirsel})}.

Our analysis revealed a qualitatively different behavior of the dynamics of phenotypic evolution for large- versus small-effect loci. In particular, we find that the mean deviation vanishes exponentially fast over a time scale 
\begin{subnumcases}
{\tau \propto}
(s \ell {\bar \gamma}^2)^{-1}~&~ ${\bar \gamma} < {\hat \gamma}$~~ \textrm{(small effects)} \label{tau1} \\
(s z_f {\bar \gamma})^{-1}  ~&~ ${\bar \gamma} > {\hat \gamma}$~~ \textrm{(large effects)} \label{tau2}~.
\end{subnumcases}
When most effects are small, the time $\tau$ depends explicitly on $\ell$ which shows that almost all loci under selection participate in the adaptation process and therefore polygenic adaptation can be rapid. These fast dynamics are due to the large initial genetic variance; in fact, the factor $\ell {\bar \gamma}^2$ in equation (\ref{tau1}) is the stationary genetic variance in the population \citep{Vladar:2014}. In contrast, when most effects are large, the initial genetic variance is small, only a few major loci play an important role over short times and the time scale of adaptation is determined by the large size of the phenotypic \tc{black}{effects}. Here, at short times, the genetic variance increases dramatically (see, Figure 4B of \citet{Jain:2017}) and the allele frequencies at several major loci undergo selective sweeps. 

\section*{Detecting rapid polygenic adaptation in the genome}

The directional selection model also allows us to predict the minimum size of the phenotypic effect required for a selective sweep to occur at a major locus. When most effects are small, our analysis shows that {an effect size larger than the initial variance is required for a large change in the allele frequency  (see (38) of \citet{Jain:2017})}; however, for exponentially distributed effects, the probability of such events is exceedingly small for large $\ell$ and therefore selective sweeps are unlikely when a phenotypic trait is controlled by many small-effect loci. \tc{black}{When most effects are large, we find that the allele frequency at a locus with an effect size larger than the mean effect may undergo a large shift (see (41) of \citet{Jain:2017}). As many loci satisfy this condition, selective sweeps occur at several major loci when many large-effect loci determine a trait.}

The classical sweeps described above can be detected by powerful methods that have been developed in the last 15 years (reviewed in \citet{Stephan:2016}). However, {in natural populations}, selective sweeps appear to be relatively rare in agreement with the previous observations \citep{Chevin:2008,Pavlidis:2012,Wollstein:2014}. On the other hand, in the case of domestication, numerous examples have been described in the literature in which selective sweeps overlap with known quantitative trait loci (QTL) in pigs, chicken and cattle \citep{Rubin:2012,Axelsson:2013, Qanbari:2014}. This may be attributed to the action of artificial selection during domestication, which causes larger optimum shifts than selection in natural populations. Indeed, under such circumstances, our criteria predict more sweeps to occur (see equation (41) of \citet{Jain:2017}).

When most effects are large, in addition to classical sweeps, occasionally we find large allele frequency shifts that resemble sweeps to some extent but are very slow and thus do not occur within the short-term phase in which the classical sweeps are predicted. Such an example can be found in Figure 3 of Jain and Stephan (2017). In this case, an allele increases from a low frequency (below $0.1$) to more than $0.9$ on a time scale that is about three orders of magnitude larger than the short-term phase. Clearly, such an allele would not lead to features that are hallmarks of selective sweeps (for instance, a strong reduction of neutral variation around the selected locus). Therefore, this allele would probably remain undetected by the available methods used to identify selective sweeps \citep{Stephan:2016}.

{Unlike selective sweeps, detecting small shifts in allele frequency at minor loci as a response to very recent selection is still a challenging problem. Standard methods that look for either large differences in allele frequency between different geographic regions \citep{Foll:2008,Riebler:2008} or a strong correlation between (a suitably transformed) trait mean and environmental variables \citep{Coop:2010,Berg:2014} and are widely used in human population genetics are not suitable when adaptation occurs rapidly \tc{black}{because the frequency gradients across geographic regions or phenotype-environment correlations} may not be sufficiently large on the short time scales over which 
rapid adaptation occurs \citep{Stephan:2016}. The same holds for other population differentiation methods that were specifically developed to detect polygenic adaptation  \citep{Turchin:2012,Robinson:2015,Racimo:2017}.}

{To our knowledge, the only method that appears to be suitable for detecting genomic signatures of rapid polygenic adaptation is a new technique that focuses on patterns of variation around each selected SNP to infer recent changes in the relative frequencies of the two alleles \citep{Field:2016}. The idea underlying this approach is that recent selection distorts the ancestral genealogy of sampled haplotypes at a selected site. The terminal branches of the genealogy tend to be shorter for the favored allele than for the other one, and hence haplotypes carrying the favored allele tend to carry fewer singletons. 
To evaluate whether one can use this method to detect signatures of recent polygenic selection, \citet{Field:2016} modified the sign of the singleton density score so that it reflects the change in frequency of the + allele of a trait (instead of the derived allele) and applied the technique to human height, a highly polygenic trait \citep{Turchin:2012,Robinson:2015}.} Based on a set of more than $550$ independent height-associated SNPs, they found evidence of an increase in human height during the past $2000$ years (corresponding to about $80$ generations) in a British population sample. Thus, this method appears to be sufficiently powerful to detect small phenotypic changes over a very short time period when the trait is highly polygenic. According to our model, the underlying reason for this observation may be that in the short-term phase, the response of the allele frequencies to an environmental change is correlated in the sense that the majority of them (if not all) shift in the same direction (as predicted by equation (\ref{dirsel})).

\section*{Summary and future directions}

In this review, we report some recent progress in understanding fast polygenic adaptation. In \citet{Jain:2015,Jain:2017}, we  studied a quantitative genetic model of adaptation with explicit reference to population genetics. Our analysis shows that fast polygenic adaptation may be caused by two qualitatively very different mechanisms: strong positive directional selection (leading to selective sweeps) at a few loci of large effects or subtle frequency shifts of alleles at many loci of small effects. Furthermore, combinations of these basic processes may also lead to rapid adaptation.

However, there are several caveats that might question the generality of our conclusions and need to be addressed in future studies. We considered only a single quantitative trait, which is controlled by a finite number of diallelic loci. Thereby, we ignored the findings of association studies that selection affecting one trait may often affect many other traits (pleiotropy) (e.g., \citet{Boyle:2017}). Furthermore, we neglected dominance and epistasis, so that the trait is determined additively. The recombination rate between loci is assumed to be high relative to selection so that there is linkage equilibrium between loci and the mutations between the two loci occur at equal rates. Based on these assumptions, the ODEs for the allele frequency changes at each locus could be formulated (cf. equation (\ref{pevoleqnf})). However, despite their relative simplicity, \tc{black}{a solution of these ODEs that is valid for all times could not be obtained analytically.} Only in the short-term phase, {\it i.e.}, in the time period in which the phenotypic mean reached a value close to the new optimum after a sudden environmental shift, the ODEs could be approximated by differential equations that take a form known for positive directional selection (but scaled by the distance from the new optimum). This generalized directional selection model could then be treated analytically.

Our analysis of this model revealed the aforementioned qualitatively different behavior of large- versus small-effect loci in the short-term phase after a sudden environmental change of the optimum. Perhaps the most interesting result of this treatment is that fast adaptation may occur for polygenic traits that are mostly determined by genes of small effects when the number of loci involved is sufficiently large (see equation (\ref{tau1})). This result appears to contradict the notion that selective sweeps are the predominant mode of fast adaptation \citep{Messer:2013} since many phenotypic traits such as human height \citep{Visscher:2008,Turchin:2012,Robinson:2015,Field:2016} may be highly polygenic. 

Relaxing the aforementioned assumptions of the model is challenging. In particular, modeling the action of selection on multiple traits appears to be difficult at present, despite the emerging literature on trait architecture in humans 
\citep{Boyle:2017}. Yet, a promising step in quantifying the effects of pleiotropy has recently been made by 
\citet{Simons:2017} who analyzed a highly pleiotropic selection model at equilibrium. For some of the other assumptions, however, it should be possible to extend the analysis in a straightforward manner. For example, the symmetry assumption that the mutation rates between alleles in both directions are equal can readily be relaxed. 

Another perhaps more important problem of our analysis is that in our model of rapid polygenic adaptation, we made the unrealistic assumption that populations are infinitely large and neglected the effects of genetic drift and demography. However, both likely play an important role for populations undergoing sudden environmental shifts. For instance, after an environmental change a small part of the population may enter a new habitat while the parental population remains in the previous environment \citep{Innan:2008}. The derived population may thereby undergo a population size reduction (bottleneck). It is therefore important to study polygenic selection for populations that are finite in size \citep{Bodova:2016,Franssen:2017} and may undergo size changes in time.

Concerning applications of the theory developed here, in the case of mostly large effects, it is possible to resort to the methods that have been developed for sweep detection in the case of individual genes such as SweepFinder \citep{Nielsen:2005b} or SweeD \citep{Pavlidis:2013}, which correct for the effects of drift and demography. However, when most effects are small, new methods need to be designed that analyze small allelic frequency shifts for populations of varying size. It is expected that under the joint action of selection and genetic drift allele frequencies at small-effect loci do not change simultaneously in the same direction (as in the deterministic model discussed above), since drift tends to drive intermediate allele frequencies toward zero or one \citep{Pavlidis:2012}. This, however, may reduce the power of tests of polygenic selection considerably and needs to be explored in detail to obtain rough estimates of the number of trait-associated SNPs required for the tests. \\

\noindent{\bf Acknowledgements:} The authors thank Brian Charlesworth and two anonymous reviewers for useful comments. KJ thanks the organisers of the symposium `Genetics of Adaptation' for an interesting meeting. The research of WS was supported by the German Research Foundation DFG (grant Ste 325/17-1 from the Priority Program 1819).


\clearpage

\clearpage

\clearpage

\clearpage


\end{document}